# Transiciones tecnológicas y los límites de la inferencia en sistemas educativos adaptativos

# Technological Transitions and the Limits of Inference in Adaptive Educational Systems

Hugo Roger Paz
PhD Professor and Researcher Faculty of Exact Sciences and Technology National University of Tucumán
Email: hpaz@herrera.unt.edu.ar
ORCID: https://orcid.org/0000-0003-1237-7983

**Resumen**
En los sistemas educativos contemporáneos, los indicadores de desempeño académico desempeñan un papel central en la evaluación institucional y en la interpretación de trayectorias estudiantiles. Sin embargo, en contextos caracterizados por una rápida transformación tecnológica, la validez inferencial de estos indicadores resulta cada vez más problemática. Este artículo analiza cómo, en sistemas educativos adaptativos, inferencias estadísticamente correctas pueden volverse sistemáticamente engañosas cuando se producen cambios estructurales en las condiciones de funcionamiento del sistema.

Desde un enfoque teórico e interpretativo, el trabajo conceptualiza las transiciones tecnológicas como perturbaciones estructurales exógenas que reconfiguran incentivos, restricciones y estrategias de participación, sin implicar necesariamente un deterioro de las capacidades subyacentes del estudiantado. A partir de evidencia empírica previa utilizada con fines ilustrativos, se examinan patrones recurrentes de inestabilidad inferencial, tales como cambios de nivel, reconfiguraciones de tendencia y aumentos de la heterogeneidad entre cohortes.

El análisis se apoya en aportes de la teoría de sistemas adaptativos complejos, la sociología de la cuantificación y la teoría de la medición para mostrar cómo la adaptación estratégica de los agentes puede desacoplar el significado de los indicadores respecto de los constructos que pretenden representar. El artículo concluye subrayando la necesidad de una interpretación cautelosa de las métricas educativas en contextos de cambio estructural.

**Palabras clave:** sistemas educativos adaptativos, transiciones tecnológicas, inferencia educativa, métricas de desempeño, deriva semántica.

**Abstract**
In contemporary educational systems, academic performance indicators play a central role in institutional evaluation and in the interpretation of student trajectories. However, under conditions of rapid technological change, the inferential validity of such indicators becomes increasingly fragile. This article examines how, in adaptive educational systems, statistically correct inferences may nevertheless become systematically misleading when structural conditions change.

Adopting a theory-informed interpretive approach, the paper conceptualises technological transitions as exogenous structural perturbations that reconfigure incentives, constraints, and participation strategies, without necessarily implying a deterioration of underlying student capabilities. Drawing on prior empirical evidence for illustrative purposes, the analysis identifies recurring patterns of inferential instability, including level shifts, trend reconfigurations, and increased heterogeneity across cohorts.

The argument integrates insights from complex adaptive systems theory, the sociology of quantification, and measurement theory to show how strategic behavioural adaptation can decouple the meaning of performance metrics from the constructs they are intended to represent. The paper concludes by emphasising the need for inferential caution when interpreting educational metrics in contexts of structural and technological transformation.

**Introducción. Cuando la inferencia correcta induce a error**

La investigación educativa se apoya cada vez más en métricas de rendimiento longitudinales para evaluar sistemas, instituciones y cohortes. Las tasas de finalización, las probabilidades de éxito, el tiempo hasta la titulación, los indicadores de compromiso y las puntuaciones de riesgo de alerta temprana se estiman habitualmente con métodos que cumplen altos estándares estadísticos. Los modelos se especifican cuidadosamente, la incertidumbre se cuantifica y se realizan comprobaciones de robustez. Desde un punto de vista técnico, las inferencias resultantes suelen ser correctas.

Sin embargo, ha surgido una tensión persistente entre la exactitud estadística y la interpretación sustantiva. En múltiples contextos educativos, tendencias empíricamente robustas —descensos en la finalización, aumentos en las tasas de no participación, cambios en el compromiso— se interpretan con frecuencia como evidencia de un deterioro de la capacidad de los estudiantes, una caída de los estándares académicos o un fracaso institucional. Estas interpretaciones suelen persistir incluso cuando no se puede establecer una degradación proporcional en la estructura curricular, los recursos didácticos o las características de ingreso.

Este artículo sostiene que tales interpretaciones erróneas no son principalmente el resultado de datos deficientes, modelos débiles o negligencia metodológica. Por el contrario, reflejan un problema más profundo de inestabilidad inferencial: una condición en la que los indicadores válidos pierden su significado estable porque el sistema que los genera ha experimentado un cambio estructural. En los sistemas sociales adaptativos, los agentes responden estratégicamente a los incentivos, las limitaciones y las tecnologías disponibles. Cuando estas condiciones cambian, la relación entre las métricas observadas y los constructos subyacentes puede desacoplarse, produciendo tendencias que son estadísticamente sólidas pero sustantivamente engañosas.

Las transiciones tecnológicas —específicamente, la difusión de plataformas digitales, la conectividad ubicua y la mediación algorítmica— se tratan aquí como choques estructurales exógenos. Es importante destacar que este trabajo no adopta una postura tecnológicamente determinista. No se postula la tecnología como un motor causal directo de los resultados educativos. En su lugar, las transiciones tecnológicas alteran la estructura de oportunidad dentro de la cual operan los agentes, remodelando las estrategias de comportamiento, la asignación temporal del esfuerzo y los patrones de compromiso. Bajo tales condiciones, las métricas establecidas pueden conservar su validez interna mientras pierden su significado externo.

El presente manuscrito se sitúa analíticamente aguas abajo de trabajos empíricos previos que documentaron cambios temporales estructurales en los indicadores de rendimiento educativo a través de eras tecnológicas. Ese trabajo estableció la existencia de cambios de nivel estadísticamente significativos, cambios de pendiente y reconfiguraciones de la varianza alineados con periodos tecnológicos exógenos, absteniéndose explícitamente de la atribución causal. Partiendo de esos hallazgos, el objetivo aquí no es cuestionar las regularidades empíricas observadas, sino interrogar el salto inferencial desde esas regularidades hacia afirmaciones sustantivas sobre el declive, la erosión de la calidad o el déficit de los estudiantes.

La contribución de este artículo es, por tanto, conceptual más que empírica. Desarrolla una explicación fundamentada teóricamente de por qué la inferencia correcta puede, no obstante, inducir a error en sistemas educativos adaptativos que atraviesan una transición tecnológica.

Reconocer este fenómeno no es un argumento en contra de la medición, sino un argumento a favor de la humildad inferencial y la conciencia estructural. En este sentido, la educación sirve como un caso paradigmático —más que como un mero ejemplo— para observar la inestabilidad inferencial. Los sistemas educativos presentan propiedades específicas que los convierten en entornos de alta ganancia para este análisis. En primer lugar, la alta capacidad adaptativa de los estudiantes permite una reconfiguración estratégica rápida en respuesta a los incentivos cambiantes. En segundo lugar, los costes de cambio entre diferentes estrategias de compromiso son relativamente bajos en comparación con otros dominios sociales. En tercer lugar, la densidad de la vigilancia métrica en la educación moderna proporciona un registro granular de la deriva del comportamiento. En cuarto lugar, y quizás lo más crítico, existe un acoplamiento débil entre las métricas de rendimiento y los resultados materiales inmediatos; las calificaciones influyen en las trayectorias a largo plazo, pero raramente determinan la supervivencia a corto plazo. Esta combinación de factores permite una mayor experimentación en las respuestas de comportamiento ante las nuevas tecnologías. Estas mismas dinámicas son epistémicamente reveladoras para otros dominios bajo mediación algorítmica, como los mercados laborales y el gobierno de plataformas. La educación, por tanto, no es única en sus desafíos; más bien, sus propiedades estructurales hacen que el problema subyacente de la deriva semántica sea únicamente visible, proporcionando una ventana clara a cómo los sistemas adaptativos se reconfiguran bajo choques estructurales.

Al sintetizar ideas de la sociología de la cuantificación, la teoría de los sistemas adaptativos complejos y la teoría de la medición, este artículo articula un problema general: las métricas pueden permanecer estadísticamente estables mientras su significado se desplaza.

**Marco teórico. Métricas, adaptación y límites inferenciales**
*Las métricas en los sistemas sociales adaptativos*

Las métricas suelen tratarse como representaciones neutrales de constructos latentes como la capacidad, el aprendizaje o el compromiso. Esta visión presupone una relación relativamente estable entre los indicadores observables y los fenómenos subyacentes. En los sistemas sociales adaptativos, sin embargo, esta presuposición es frágil.

Un cuerpo sustancial de investigación en la sociología de la cuantificación ha demostrado que las métricas no son meramente descriptivas, sino performativas: moldean activamente el comportamiento, las prioridades organizativas y los procesos de toma de decisiones

(Espeland & Stevens, 1998; Power, 1997). Una vez integrados en los regímenes de gobernanza, los indicadores dejan de funcionar como espejos pasivos de la realidad y se convierten en componentes del sistema que miden.

Esta idea se formaliza en dos principios bien conocidos. La Ley de Goodhart establece que cuando una medida se convierte en un objetivo, deja de ser una buena medida (Goodhart, 1975). La Ley de Campbell advierte que cuanto más se utiliza un indicador social cuantitativo para la toma de decisiones, más sujeto está a las presiones de corrupción y más distorsiona los procesos que pretende supervisar (Campbell, 1979). Aunque a menudo se invocan en discusiones sobre el fraude o la manipulación, estas leyes describen una propiedad estructural más general: en los sistemas adaptativos, los agentes responden estratégicamente a las métricas, no pasivamente.

Los sistemas educativos ejemplifican esta dinámica. Los estudiantes, los instructores y las instituciones aprenden qué se recompensa, qué es costoso y qué se puede posponer. Los indicadores de rendimiento agregan, por tanto, no solo la capacidad subyacente, sino también las respuestas estratégicas a las estructuras de incentivos. Crucialmente, esta agregación puede permanecer estadísticamente regular incluso cuando su contenido semántico cambia.

### *Estabilidad, validez y deriva semántica*

Las discusiones tradicionales sobre la medición se centran en la fiabilidad y la validez. La fiabilidad se refiere a la consistencia; la validez se refiere a si un indicador mide lo que afirma medir (Messick, 1989). En contextos longitudinales, sin embargo, una tercera dimensión resulta crítica: la estabilidad semántica. Un indicador puede seguir siendo fiable e internamente válido mientras su significado cambia porque el comportamiento que agrega ha sido reconfigurado.

Este fenómeno está estrechamente relacionado con lo que los teóricos de la medición describen como deriva del constructo: el cambio gradual o abrupto en el significado de un constructo a lo largo del tiempo (Haertel, 2006). La deriva del constructo puede ocurrir debido a una reforma curricular, un cambio cultural o, como se sostiene aquí, una transición tecnológica. Cuando los constructos se desplazan, la comparabilidad longitudinal se vuelve problemática incluso si los modelos estadísticos siguen estando bien especificados.

La investigación sobre la invarianza de la medición ha demostrado repetidamente que los indicadores pueden fallar al mantener la equivalencia a través de contextos, poblaciones o modos de administración (Vandenberg & Lance, 2000). Es importante destacar que el fallo de la invarianza no se manifiesta necesariamente como datos ruidosos o un mal ajuste del modelo. En su lugar, puede producir tendencias fluidas e interpretables que, no obstante, están desalineadas con los constructos que los investigadores creen estar rastreando.

En los sistemas educativos adaptativos, la deriva semántica se ve exacerbada por el comportamiento estratégico. Indicadores como el tiempo hasta la titulación, la finalización de cursos o las tasas de no participación suponen implícitamente regímenes de comportamiento estables. Cuando esos regímenes cambian —porque los costes de oportunidad se desplazan, surgen nuevas credenciales o se reconfiguran las estrategias de compromiso— la misma métrica puede pasar a indexar un fenómeno subyacente diferente.

### *Transiciones tecnológicas y reconfiguración estructural en educación*

Este artículo conceptualiza las transiciones tecnológicas como choques estructurales en el proceso de generación de datos de los sistemas educativos. El término "choque" no implica una interrupción repentina o una primacía causal. Más bien, denota una alteración exógena de las limitaciones y las asequibilidades (affordances) que redefine el espacio de estrategia disponible para los agentes.

Desde la perspectiva de los sistemas adaptativos complejos, tales choques alteran las condiciones de contorno más que las reglas internas (Holland, 1992; Cilliers, 1998). Los agentes siguen orientados a objetivos y son adaptativos, pero el entorno en el que ocurre la adaptación cambia. Las nuevas tecnologías reducen el coste marginal del acceso a la información, aumentan la flexibilidad en el ritmo y la secuencia, e introducen mecanismos alternativos de señalización. Estos cambios remodelan la forma en que los agentes racionales asignan tiempo, esfuerzo y atención.

Una implicación crítica es que los impactos tecnológicos rara vez son sincrónicos. La investigación profunda en múltiples contextos educativos indica que la adopción, la integración y la adaptación del comportamiento son desiguales y están rezagadas. Las instituciones, las disciplinas y las poblaciones de estudiantes responden a diferentes velocidades. Como resultado, las comparaciones entre cohortes e instituciones que ignoran el desajuste temporal corren el riesgo de confundir el tiempo de adaptación con diferencias de rendimiento.

Desde un punto de vista inferencial, esta asincronía socava la suposición de que las tendencias observadas reflejan un cambio homogéneo en la capacidad subyacente. En su lugar, pueden reflejar una adaptación escalonada a un paisaje de limitaciones cambiante.

### *La Crítica de Lucas y la dependencia del régimen*

El desafío inferencial planteado por la transición tecnológica está estrechamente alineado con la Crítica de Lucas, articulada originalmente en macroeconomía (Lucas, 1976). La crítica sostiene que no se puede suponer que las relaciones estimadas bajo un régimen de política se mantengan bajo otro, porque los agentes ajustan su comportamiento en respuesta al propio régimen.

Aplicado a la educación, la implicación es directa: los parámetros estimados bajo un régimen pre-digital no pueden presumirse invariantes bajo un régimen digital o mediado algorítmicamente. Cuando las reglas de interacción cambian —a través de la plataformización, la retroalimentación en tiempo real o nuevas vías de acreditación— se redefinen las relaciones estructurales que vinculan el esfuerzo, el compromiso y los resultados.

Esta dependencia del régimen no implica que la inferencia se vuelva imposible. Implica que la inferencia se vuelve condicional. Los modelos pueden seguir ajustándose bien a los datos mientras codifican relaciones que ya no corresponden a constructos estables. Sin el reconocimiento explícito del cambio de régimen, los analistas corren el riesgo de confundir la reconfiguración estructural con el deterioro.

### *Límites inferenciales en sistemas educativos adaptativos*

En conjunto, la naturaleza performativa de las métricas, la propensión a la deriva del constructo y la dependencia del régimen de las relaciones de comportamiento definen un límite fundamental de la inferencia en los sistemas educativos adaptativos. Los indicadores observados combinan múltiples componentes: capacidad subyacente, adaptación estratégica, cumplimiento de los regímenes de medición y respuestas a las asequibilidades tecnológicas.

Distinguir entre estos componentes es un reto estadístico, particularmente cuando la adaptación ocurre de forma gradual y heterogénea. En tales contextos, la confianza inferencial puede aumentar paradójicamente en el mismo momento en que la interpretación sustantiva se vuelve más frágil.

Este trabajo no sostiene que las métricas sean inútiles, ni que el análisis cuantitativo deba abandonarse. Sostiene que la exactitud estadística es insuficiente cuando las condiciones estructurales cambian. La validez inferencial requiere atención no solo a los modelos y los datos, sino a la relación evolutiva entre los indicadores y el significado.

## Las transiciones tecnológicas como choques estructurales en los sistemas educativos
### Cambio tecnológico sin determinismo tecnológico

El papel de la tecnología en el cambio educativo se ha enmarcado a menudo en términos deterministas, oscilando entre promesas utópicas de eficiencia y narrativas distópicas de degradación cognitiva. Ambas posiciones oscurecen la perspectiva analíticamente más productiva: la tecnología como un modificador estructural de los espacios de acción más que como un motor causal directo de los resultados (Selwyn, 2016; Biesta, 2010).

Desde una perspectiva de sistemas, las transiciones tecnológicas operan alterando las limitaciones, las asequibilidades y los costes de oportunidad. Modifican cómo se organizan las actividades educativas en el tiempo y el espacio, cómo se accede a la información y se procesa, y cómo se distribuye el esfuerzo entre demandas competidoras. Es importante destacar que estos cambios no afectan directamente a la capacidad cognitiva o al contenido curricular. En su lugar, remodelan el entorno en el cual los agentes —estudiantes, instructores, instituciones— toman decisiones estratégicas.

En los sistemas adaptativos complejos, tales cambios ambientales funcionan como choques estructurales exógenos. Redefinen el conjunto de estrategias factibles sin prescribir comportamientos específicos (Holland, 1992; Arthur, 1999). Los agentes responden de forma heterogénea, experimentando con nuevos patrones de compromiso, ritmo y señalización. Los resultados agregados emergen de estas adaptaciones locales más que de un cambio de comportamiento uniforme.

Esta distinción es crucial para la inferencia. Si las transiciones tecnológicas reconfiguran principalmente el comportamiento más que la capacidad, entonces las métricas sensibles a los patrones de comportamiento pueden registrar un cambio sistemático incluso cuando la capacidad subyacente permanece estable. Interpretar ese cambio como declive confunde la adaptación con el deterioro.

### Difusión asincrónica y adaptación rezagada

Un hallazgo consistente en la literatura de investigación profunda es que las transiciones tecnológicas no son ni instantáneas ni sincrónicas. Las curvas de adopción varían entre instituciones, disciplinas, grupos socioeconómicos y contextos nacionales (Rogers, 2003; Selwyn, 2019). Incluso cuando el acceso es generalizado, la integración en las prácticas cotidianas es desigual y está temporalmente rezagada.

Esta asincronía tiene implicaciones directas para la inferencia longitudinal. Cuando las cohortes experimentan diferentes regímenes tecnológicos en diferentes etapas de sus trayectorias educativas, los resultados observados reflejan una mezcla de comportamientos previos a la transición, de transición y posteriores a la misma. Por lo tanto, las rupturas de tendencia pueden parecer graduales, retrasadas o desalineadas temporalmente con la introducción nominal de la tecnología.

En tales contextos, las suposiciones convencionales de homogeneidad temporal se vuelven insostenibles. La ausencia de una discontinuidad brusca no implica continuidad del significado. Por el contrario, a menudo señala una adaptación escalonada que ocurre bajo tendencias agregadas fluidas.

**Señales empíricas de ruptura inferencial**
*Posicionamiento de la evidencia empírica*

La evidencia empírica discutida en esta sección no se presenta como un nuevo análisis. Más bien, cumple una función ilustrativa e interpretativa, fundamentando el argumento conceptual en regularidades estructurales documentadas. Los análisis subyacentes establecieron reconfiguraciones temporales estadísticamente significativas en indicadores clave de rendimiento educativo a través de eras tecnológicas, evitando explícitamente la atribución causal.

*Cambios de nivel estructurales. líneas de base reconfiguradas*

A través de múltiples cursos troncales y cohortes, el análisis de puntos de ruptura estructurales identificó cambios de nivel significativos en las probabilidades de éxito y las tasas de no participación alineados con periodos tecnológicos amplios. Los regímenes iniciales se caracterizaron por mayores probabilidades medias de éxito y menores tasas de resultados de 'suspenso sin intento'. Los periodos de transición mostraron patrones mixtos, mientras que los regímenes posteriores mostraron una probabilidad media de éxito sistemáticamente menor acompañada de tasas elevadas de no participación.

Crucialmente, el aumento de los resultados de falta de intento (no-attempt) justifica un replanteamiento como una respuesta adaptativa de contorno más que como un indicador de déficit. La no participación —específicamente la elección de no presentar trabajos evaluables a pesar de permanecer matriculado— representa una desincorporación estratégica altamente sensible a la flexibilidad proporcionada por los nuevos regímenes tecnológicos. Desde una perspectiva de sistemas adaptativos, este comportamiento señala estructuras de incentivos desalineadas donde los agentes redirigen el esfuerzo hacia actividades con mayores retornos percibidos. Esto contrasta con las narrativas de déficit que enmarcan la falta de intento como debilidad académica o fallo motivacional. A diferencia de los constructos latentes, la no participación es un comportamiento observable que refleja cómo los agentes optimizan el compromiso cuando los costes del aplazamiento disminuyen. En entornos de alta flexibilidad, la falta de intento refleja una reasignación racional más que una incapacidad. Al considerar esto como una respuesta adaptativa legítima, los analistas evitan patologizar cambios de comportamiento consistentes con la transición estructural.

Desde un punto de vista puramente descriptivo, estos cambios son inequívocos. El problema inferencial surge en su interpretación. Una lectura ingenua sugiere un deterioro en el rendimiento o la preparación de los estudiantes. Sin embargo, varias características de los datos socavan esta conclusión.

En primer lugar, los cambios de nivel son consistentes entre cursos, apareciendo tanto en Matemáticas como en Física a pesar de las diferencias sustanciales en el contenido y la estructura de la evaluación. Esta consistencia apunta lejos de un fallo instruccional específico de la disciplina y hacia un mecanismo a nivel de sistema.

### *Reconfiguraciones de tendencia. cambios de pendiente y puntos de inflexión*

El análisis de regresión segmentada reveló cambios sistemáticos de pendiente a través de los regímenes tecnológicos. En los periodos iniciales, las probabilidades de éxito eran planas o crecientes. Los periodos de transición exhibieron puntos de inflexión claros, con tendencias que se invirtieron hacia pendientes negativas. Los periodos posteriores mostraron estabilización, con pendientes próximas a cero.

Este patrón es particularmente revelador. Si el sistema estuviera experimentando un deterioro continuo de la capacidad, se esperaría un declive monotónico. En su lugar, los datos muestran una reconfiguración seguida de una estabilización. Tales dinámicas son características de sistemas adaptativos que responden a nuevas limitaciones más que de sistemas sometidos a una degradación continua (Arthur, 1999).

### *Inflación de la varianza e inestabilidad de la cohorte*

Quizás la señal empírica más subestimada es el cambio en la estructura de la varianza. El análisis de la varianza indica un aumento de la variabilidad entre cohortes durante y después de la transición tecnológica, particularmente para indicadores de comportamiento como las tasas de no participación.

En la teoría clásica de la medición, el aumento de la varianza se interpreta a menudo como ruido. En los sistemas adaptativos, sin embargo, la varianza puede señalar una adaptación heterogénea. Diferentes cohortes, expuestas a condiciones estructurales similares, experimentan con diferentes estrategias. Con el tiempo, la varianza puede estabilizarse en un nivel más alto a medida que coexisten múltiples equilibrios de comportamiento.

## **Modos de fallo inferencial bajo cambio estructural**

### *Malinterpretar la adaptación como declive*

El primer y más generalizado fallo inferencial es la malinterpretación del comportamiento adaptativo como declive. Cuando los agentes responden al aumento de la flexibilidad aplazando la finalización, reasignando el esfuerzo o desconectándose selectivamente de evaluaciones de baja recompensa, las métricas sensibles a la regularidad y al ritmo registran un deterioro.

Esta lectura errónea es especialmente pronunciada en la interpretación de la no participación como un síntoma de desconexión-como-fracaso. Los marcos dominantes suelen situar la falta de intento junto al absentismo y el riesgo de abandono, asumiendo que el hecho de no participar refleja una incapacidad fundamental para cumplir los estándares académicos. Sin embargo, cuando los regímenes tecnológicos expanden la flexibilidad temporal —facilitada por el acceso ubicuo y las vías asincrónicas— la no participación puede funcionar como una gestión estratégica de los contornos. Los estudiantes pueden mantener la matrícula formal para preservar el acceso institucional o la elegibilidad financiera, mientras aplazan el compromiso real hasta que su paisaje individual de costes de oportunidad sea más favorable. El error inferencial reside en atribuir esta reconfiguración del comportamiento a déficits latentes de los estudiantes sin descartar primero las asequibilidades estructurales que hacen de esta desconexión selectiva una vía viable y racional. Reformular la no participación como una señal de desalineación sistémica en lugar de un fracaso individual permite una interpretación que respeta la capacidad agencial del estudiante al tiempo que reconoce el profundo impacto de los cambios tecnológicos en la estabilidad de las métricas educativas.

Desde la perspectiva del agente, tal comportamiento puede ser totalmente racional. Desde la perspectiva del analista, parece patológico solo si se asume un régimen de comportamiento estático. Esta suposición rara vez se hace explícitamente, pero sustenta muchas interpretaciones orientadas al déficit.

## Implicaciones de la inestabilidad inferencial

El argumento desarrollado hasta ahora ha demostrado que la inestabilidad inferencial no es un problema incidental derivado de datos ruidosos o de un descuido metodológico, sino una propiedad estructural de los sistemas educativos adaptativos que atraviesan una transición tecnológica. Esta sección articula las implicaciones de esta afirmación para tres dominios interconectados: la investigación educativa, las analíticas de aprendizaje y la interpretación de políticas. Es importante destacar que estas implicaciones son analíticas más que prescriptivas. El objetivo es aclarar cómo debe reformularse la inferencia en condiciones de cambio estructural, no abogar por intervenciones específicas.

### *Implicaciones para la investigación educativa*

La investigación educativa se ha apoyado durante mucho tiempo en indicadores longitudinales para identificar tendencias, evaluar reformas y diagnosticar el rendimiento del sistema. Los hallazgos presentados aquí cuestionan una suposición central que subyace a esta práctica: que los indicadores mantienen un significado estable a lo largo del tiempo a menos que sean invalidados explícitamente por el error de medición.

Desde un punto de vista inferencial, estos desafíos sugieren la necesidad de conceptualizar métricas con mayor resiliencia a la deriva semántica. La resiliencia métrica puede entenderse como una propiedad analítica en lugar de una prescripción de diseño: indicadores que exhiben una sensibilidad reducida a la adaptación estratégica del comportamiento y mantienen una mayor robustez a través de los cambios de régimen al priorizar la triangulación sobre la optimización de una única métrica. Como advierten tanto la Ley de Goodhart como la Ley de Campbell, los regímenes de indicador único invitan inevitablemente a la distorsión estratégica a medida que los agentes ajustan sus estrategias locales a la presión de la medición. Desde la lógica de los sistemas adaptativos complejos, la resiliencia implica que los indicadores permanecen informativamente estables incluso cuando los agentes realizan ajustes locales. Esto sugiere la necesidad de sistemas de medición que distribuyan la observación a través de múltiples dimensiones no sustituibles que se degraden con elegancia en lugar de catastróficamente bajo la adaptación. La robustez inferencial podría encontrarse en el desacoplamiento estructural de los indicadores respecto a objetivos de incentivos directos y de alto riesgo, preservando así su valor informativo a largo plazo como representaciones de los constructos subyacentes. Importante: la resiliencia no es una característica estática sino condicional, que depende de la velocidad de la reconfiguración del comportamiento dentro del sistema. Este lente analítico desplaza el foco desde la precisión técnica de la medición hacia su durabilidad como señal bajo condiciones donde el proceso generador de datos se reconfigura fundamentalmente. Al enmarcar la resiliencia en estos términos, los investigadores pueden evaluar qué indicadores tienen probabilidades de conservar su significado bajo un cambio estructural, en lugar de asumir que la corrección estadística garantiza la estabilidad semántica.

*Lo que se puede y no se puede inferir*
El análisis precedente motiva una aclaración crítica: ¿qué se puede inferir exactamente de las métricas educativas bajo condiciones de transición tecnológica?

**Conclusión. Humildad inferencial en sistemas educativos adaptativos**
Este artículo ha sostenido que, en los sistemas educativos adaptativos que atraviesan una transición tecnológica, la inferencia estadísticamente correcta puede, no obstante, inducir a error. El problema no es una estimación defectuosa, datos inadecuados o un análisis descuidado. Es estructural.

Las transiciones tecnológicas actúan como choques exógenos que reconfiguran las limitaciones, los incentivos y las estrategias de comportamiento. Los agentes se adaptan racionalmente, produciendo patrones agregados que permanecen estadísticamente regulares mientras su significado se desplaza. Las métricas conservan la validez interna pero pierden la estabilidad externa. Bajo estas condiciones, la inferencia se vuelve condicional y la interpretación se vuelve azarosa.

La educación ha servido aquí como un caso paradigmático. El argumento, sin embargo, se extiende más allá de la educación. Cualquier dominio en el que agentes adaptativos respondan a las métricas bajo limitaciones cambiantes —mercados laborales, rendimiento organizativo, gobernanza algorítmica— se enfrenta a límites inferenciales similares.

La contribución central de este manuscrito es articular este problema de forma explícita y situarlo dentro de un marco teórico coherente basado en los sistemas adaptativos complejos, la teoría de la medición y la sociología de la cuantificación. Al hacerlo, desplaza el foco de atención de la mejora de las métricas a la comprensión de los sistemas que las generan.

La implicación no es abandonar la medición, sino abordarla con humildad inferencial. Las métricas son herramientas indispensables, pero no son ventanas transparentes a la realidad. Son componentes de sistemas adaptativos, entrelazados con los comportamientos que miden.

La inferencia correcta es necesaria. No es suficiente.

**Referencias**

Arthur, W. B. (1999). Complexity and the economy. Science, 284(5411), 107–109. https://doi.org/10.1126/science.284.5411.107

Biesta, G. (2010). Good education in an age of measurement. Paradigm.

Campbell, D. T. (1979). Assessing the impact of planned social change. Evaluation and Program Planning, 2(1), 67–90. https://doi.org/10.1016/0149-7189(79)90048-X

Cilliers, P. (1998). Complexity and postmodernism: Understanding complex systems. Routledge.

Espeland, W. N., & Stevens, M. L. (1998). Commensuration as a social process. Annual Review of Sociology, 24, 313–343.

Fisher, A., Rudin, C., & Dominici, F. (2019). All models are wrong, but many are useful: Learning a variable's importance by studying an entire class of prediction models simultaneously. Journal of Machine Learning Research, 20(177), 1–81.

Goodhart, C. A. E. (1975). Problems of monetary management: The UK experience. En Papers in Monetary Economics (Vol. 1). Reserve Bank of Australia.

Haertel, E. H. (2006). Reliability. En R. L. Brennan (Ed.), Educational measurement (4.ª ed., pp. 65–110). Praeger.

Holland, J. H. (1992). Adaptation in natural and artificial systems. MIT Press.

Lucas, R. E. (1976). Econometric policy evaluation: A critique. En K. Brunner & A. H. Meltzer (Eds.), The Phillips Curve and labor markets (pp. 19–46). North-Holland.

Messick, S. (1989). Validity. En R. L. Linn (Ed.), Educational measurement (3.ª ed., pp. 13–103). Macmillan.

Power, M. (1997). The audit society: Rituals of verification. Oxford University Press.

Rogers, E. M. (2003). Diffusion of innovations (5.ª ed.). Free Press.

Selwyn, N. (2016). Education and technology: Key issues and debates. Bloomsbury.

Selwyn, N. (2019). Should robots replace teachers? British Journal of Educational Technology, 50(4), 1627–1639. https://doi.org/10.1111/bjet.12771

Vandenberg, R. J., & Lance, C. E. (2000). A review and synthesis of the measurement invariance literature. Organizational Research Methods, 3(1), 4–70. https://doi.org/10.1177/109442810031002